\documentclass{elsart}

\usepackage{natbib}

\usepackage{graphicx}
\usepackage{amssymb}

\def\o {\ensuremath{^{\circ}}}

\begin{document}

\begin{frontmatter}



\title{Crown Detectors Arrays to Observe Horizontal and Upward Air-Showers}

 \author[label1,label2]{D. Fargion}
 \author[label1]{, M. Grossi}
  \author[label1]{, M. De Santis}
  \author[label1]{ P.G. De Sanctis Lucentini}
  \author[label1,label2]{, M.Iori}
  \author[label1]{, A. Sergi}
   \author[label1]{, F.Moscato}
% \address[label1]{Dep. of Physics, University of Rome "La Sapienza"}
\address[label1]{Physics Department , INFN, Universit\'a di Roma  "La Sapienza", Pl. A. Moro
2, 00185, Rome,  Italy}
\address[label2]{ INFN, Universit\'a di Roma1, "La Sapienza"}

%\author[label1]{M. Grossi}

%\address{}

\begin{abstract}
 Terrestrial Cerenkov Telescopes at tens GeV gamma energy  and Scintillators set on a
 Crown-like  array facing the Horizons may reveal
far  Cosmic Rays Showers or  nearer PeVs Neutrino
$\overline{\nu}_{e}-e \rightarrow W^-$ shower in air as well as
upgoing ${\nu}_{\tau} + N \rightarrow \tau + X $,$\tau\rightarrow
$ Earth-Skimming tau air-showers. Even UHE SUSY $\chi^o +
e\rightarrow \tilde{e} \rightarrow\chi^o + e $ at tens PeVs-EeV
energy may blaze at Horizons, as $\overline{\nu}_{e}-e
\rightarrow W^-$ shower.
 We show first estimate  on down and up-going Horizontal Showers
 traces for present and future Magic-like  Crown Arrays and their  correlated Scintillator-like  twin Crown Arrays.
  The one mono or stereo-Magic   elements facing the Horizons  are already comparable to  present
  Amanda underground  neutrino detector.

\end{abstract}


\end{frontmatter}

% main text
\section{Horizons: A new Frontier of High Energy Astrophysics}

Downward  but inclined  or horizontal  Cosmic Ray (C.R.) Showers
($70^o-90^o$ zenith angle) produce secondary ($\gamma$,$e ^\pm$)
mostly suppressed by high  atmosphere column  depth. The Upward
Air-Showers are totally
 absorbed by the Earth  shadows and in principle are free from any background.
 Most inclined down-ward C.R. air-shower  produce in high altitudes
 Cherenkov  photons that are diluted by the large distances and by air opacity,
  while their secondary penetrating, $\mu^\pm$  and their successive
  decay into $e^\pm$,$\gamma$, may revive the optical signal by additional Cerenkov lights.
   Tiles of scintillators (see also \cite{Iori04}), set in a circular crown array  facing the Horizons may be able to
   inspect a wide solid angle, recording the muon tracks,
    as well as their eventual electromagnetic companion bundles.
  An hybrid Crown  made by a circular  Cerenkov light Telescopes and
  by  Muon-gamma-electron scintillator arrays, may be an ideal
  detector for horizontal Cosmic Rays and Tau Air-Shower signals.
  A few arrays of such Crown Detectors located at a few kilometer one from the other
   may even better probe the  horizontal air-shower arrival,
   structure and lateral distribution.

    The larger horizontal distances widen the shower's cone into wider areas, while the
geo-magnetic field open the positive-negative charges  in a very
characteristic  fan-like shape, polarized by, and, aligned
orthogonal to, the local geo-magnetic field vector.
 These elongated inclined showers are thin, collimated , fork-shaped and diluted in
elongated areas; because of it they occur in proportion more
frequently, up to two-three order of magnitude (respect to near
but dense vertical showers). The GeVs $\gamma$ telescopes at the
top of the mountains or in Space may detect at the horizons both
PeVs up to EeV (or more energetic) hadronic cosmic rays
secondaries. Correlated details on arrival angle and crossed
column depths, the shower shape, its timing signature and photon
flash intensity, may inform us on the altitude interaction,
primary UHECR composition, lateral dispersion, off-axis geometry
and total shower energy. Additional muon-gamma Crown detectors may
capture various leptons, $\pm \mu$, and  their decayed $\pm e$ and
$\gamma$ tracks  better defining the shower composition and
development.

 %%%% addendum 1%%%%
Horizontal muon secondaries  at high energy (hundred TeV) leading
to  muon catastrophic electromagnetic shower in air horizons are
 also leading to penetrating electromagnetic air-showers: their rate at a hundred
TeVs  is  $\Phi_{\mu}(E_{\mu} = 10^{14} eV) \simeq  3-5 \cdot
10^{-12} cm^{-2} sr^{-1} s^{-1}$ is comparable to up-going
atmospheric muon (at GeVs energy) induced by atmospheric neutrino;
however they they are $not$ deflected and are well $above$ the
horizons; their consequent bright Cerenkov ring cannot compete
with the higher energy searched up-going neutrino signals. At PeVs
energies even prompt muons (by the decay of charmed hadrons) are
leading to a limited flux $3-5 \cdot 10^{-15} cm^{-2} sr^{-1}
s^{-1}$ (see ref. fig.1 \cite{Sine2001}) even smaller than
expected by muon secondaries of upgoing tau air-showers.
 Therefore we shall neglect the atmospheric up-going muon showering
 out of the competitive astrophysical UHE muon and  tau
 originated by PeVs-EeVs neutrinos in Earth.
%%%% end addendum 1%%%%

 Indeed below the horizons,
at zenith angle ($90^o-99^o$) among quite rare "single" albedo
muons, more rare up-going showers traced by muon
($e^\pm$,$\gamma$) bundles would give evidence of rare Horizontal
Tau Air-Showers (HorTaus or Earth-Skimming neutrinos), originated
by ${\nu}_{\tau}$, $\overline{\nu}_{\tau}$, (${\nu}_{\tau} + N
\rightarrow \tau + X $, $\tau\rightarrow $ hadrons and/or
electromagnetic shower around EeVs energies. See
\cite{Fargion1999},\cite{Fargion
2002a},\cite{Bertou2002},\cite{Feng2002},\cite{Tseng03}; their
 energy losses in matter has been studied for underground
 detectors in first approximations \cite{Fargion1999},\cite{Fargion 2002a},
  and more in details \cite{Fargion2004}, \cite{Jones04},
\cite{Yoshida2004}, \cite{Fargion2004b}, \cite{Fargion 2004c}.

 Their rate may be
comparable or competitive with $6.3$ PeVs $\overline{\nu}_{e}-e$
neutrino (by $W^-$ resonance at Glashow peak \cite{Gandhi98})
induced air-shower (mostly hadronic) originated above and also
below horizons, in interposed atmosphere. Finally more exotic
additional and complementary UHE SUSY $\chi^o + e\rightarrow
\tilde{e} \rightarrow\chi^o + e $ \cite{Datta04} at tens PeVs-EeV
energy may blaze, as $\overline{\nu}_{e}-e \rightarrow W^-$
shower,(above and also below horizons) by their characteristic
electromagnetic signature. These UHE $\chi^o $ are expected in
topological defect scenarios for UHECRs. Their secondary shower
blazing Cerenkov lights and distances  might be disentangled from
UHECR by Stereoscopic Telescopes such as the Magic ones or Hess,
Cangoroo array experiment. The horizontal detection sensitivity of
Magic in the present set up (if  totally devoted to the Horizons
Shower search) maybe already be comparable to AMANDA underground
neutrino detector at PeVs energies.  An array of Crown detectors
made of tiles of scintillators facing the horizons would be also
able to observe from a Mountain, or a Balloon or a satellite at
the Horizons making Cerenkov-Muons-Gamma Crown Arrays the ideal
telescope for UHECR and UHE Neutrino Astronomy \cite{Fargion
2004c}.
\subsection{Crown Arrays for Horizontal edges}
Ultra High Energy Cosmic Rays (UHECR) Showers (PeVs - EeVs) born
at high altitude in the atmosphere may be detected by telescopes
such as Magic provided that they are pointed towards the horizon.
 These Cerenkov telescopes, set in a crown circular array toward the Horizons, may be correlated by an analogous
  scintillator one able to trace muons and electromagnetic shower
  particles.  The earliest gamma and Cerenkov lights produced
   by any down-ward horizontal, zenith angle $(85^o-90^o$ ),
   cosmic rays (CR), observed by a high  ($> 2$ km) mountain
 (whose C.R. nature is mainly hadronic), while  crossing  the atmosphere
  are severely absorbed by the thick horizontal air column depth ($10^4 - 5 \cdot 10^4$ $ g \cdot
cm^{-2})$. However the most energetic C.R. blazing Cerenkov shower
 must survive and also revive during their propagation. Indeed
one has to expect that these CR showers contain also a diluted but
more penetrating component made of muon bundles
\cite{NEVOD},\cite{Decor}, which can be detected in two ways: a)
by their Cerenkov lights emitted after they decay into electrons
nearby the Telescope; b) while the muons hit into the Telescope,
blazing  ring or arc of the same muon Cerenkov lights. The latter
muons  are less abundant compared to the early peak gamma photons
produced in the shower (roughly $10^{-3}$ times lower) but more
penetrating. They are produced at high altitudes and at an
horizontal distance of $100-500$ km from the observer (placed at
2.2km above the sea level,and assuming a zenith  angle of
$85^{\circ}- 91.5^{\circ}$). Therefore these hard muon bundles
(hundred GeV) might spread in large areas of tens - hundred $km^2$
while they travel towards the observers. They are partially bent
by the local geomagnetic field, (in azimuth dependence) and they
are  aligned along an axis orthogonal to the field and they decay
into electrons producing optical Cerenkov flashes. The
geo-magnetic spread of the shower leads to a aligned Cerenkov
blaze whose shape (a thin elliptic splitted twin-lightening
shower) may probe the magnetic field and the CR origination.

%%%%%%% addendum2%%%%%%%%%%%%
One of the characteristic consequence of this geomagnetic
splitting is the twin Cerenkov lights and dots appearing in the
Magic telescope as a very unique signature of the  bending
trajectories. The estimated angle separation range in $0.5^o -
3^o$ a value that is within angular resolution of Cherenkov
telescopes. This may be the imprint and the trigger signal of most
horizontal air-showers.
%%%%%%%%%%%%addendum2%%%%%%%%
 Such a characteristic signature may be detected by
the largest gamma telescope arrays as Hess, Veritas, or better by
forthcoming stereoscopic version of Magic. We argue that their
Cerenkov flashes, either single or clustered, must take place, at
detection threshold, at least tens or hundreds times a night for
Magic-like Telescope pointing toward horizons at a zenith angle
between $85^{\o}$ and $90^{\o}$. Their easy "guaranteed" discover
may offer, we hope,  a new gauge and "meter" in CR and UHECR
detection. Their primary hadronic  shower secondaries might be
hidden  by the distance (but not their Cherenkov flashes); the
late shower tail may arise in a new form by  its secondary
muon-electron-Cerenkov chain of the electromagnetic showering.
 Moreover    a rarer   but  more exciting PeV - EeV
Neutrino ${\nu_{\tau}}$ Astronomy (whose flux is suppressed by
three-four orders of magnitude respect CR one) may arise
\emph{below the horizon} with the Earth-Skimming Horizontal Tau
Air-Showers (HorTaus) \cite{Fargion1999},\cite{Fargion 2002a};
below the horizons there are not other shower at all. These UHE
Taus are produced   inside the Earth Crust by the primary UHE
incoming neutrino   ${\nu}_{\tau}$, $\overline{\nu}_{\tau}$, and
they are generated mainly by the muon-tau neutrino oscillations
from galactic or cosmic sources, \cite{Fargion 2002a},
\cite{Fargion03}, \cite{Fargion2004}. Finally we expect also that
just  above or below the horizon edge, (within a distance of a few
hundreds of km), air-showers due to the fine-tuned
$\overline{\nu}_e$-$e\rightarrow W^-\rightarrow X$ Glashow
\cite{Gandhi98} resonance at $6.3$ PeV might  be detectable. The W
main hadronic ($2/3$) or leptonic and electromagnetic ($1/3$)
signatures  may be well observed and   their rate may be used to
calibrate a new horizontal neutrino multi-flavor neutrino
Astronomy \cite{Fargion 2002a}. Again we argue  that such a
signature of nearby nature (respect to most far away ones at same
zenith angle of hadronic nature) would be better revealed by  a
Stereoscopic twin telescope such as Magic or a Telescope array
like Hess, Veritas.
       As mentioned above additional Horizontal flashes  might arise
    by Cosmic UHE $\chi_o + e \rightarrow  \widetilde{e}\rightarrow \chi_o +
    e$ electromagnetic showers  within most SUSY models, if UHECR are born in topological
   defect decay or in their annihilation, containing a relevant component of SUSY
   particles. The UHE $\chi_o + e \rightarrow  \widetilde{e}\rightarrow \chi_o +
    e$ behaves (for light $\widetilde{e}$ masses around Z boson ones)
    as the Glashow \cite{Gandhi98} resonance peak \cite{Datta04}.
The total amount of air inspected within the characteristic field
of view of  MAGIC ($1^{\o} \cdot 2^{\o}$) at the  horizon ($360$
km.) corresponds to a (water equivalent) volume-mass larger than $
V_w\simeq 1 km^3$. However their solid angle detectable beamed
volume corresponds  to a  narrower beamed volume   $ V_w\simeq
1.36 \cdot 10^{-2}$ $km^3$,  yet  comparable to the  present
AMANDA confident volume  (for Pevs
$\overline{\nu}_e$-$e\rightarrow W^-\rightarrow X$   and EeVs
${\nu}_{\tau}$, $\overline{\nu}_{\tau} + N \rightarrow
\tau\rightarrow $ showers). Greater volumes  and masses (up to $75
km^3$) are estimated for upward  EeV tau events \cite{Fargion2005}
making Magic the most sensitive UHE neutrino detector today.

\section{ Cerenkov Flashes at Horizons by Showers and Muons}

  The ultrahigh energy cosmic rays (UHECR) have been studied
   in the past mainly through their secondary particles ($\gamma$, $e^\pm$, $\mu^\pm$)
  collected vertically in large  array detectors on the ground.
   UHECRs are rare events, however the multiple cascades occurring at high
  altitudes where the shower usually takes place, expand and amplify  the signal detectable on the ground.
  On the other hand, at the horizon the UHECR  are hardly observable (but also rarely
  searched).  They are diluted both by  the larger  distance they have to cover and
by the atmosphere opacity suppressing exponentially their
electromagnetic  secondaries (electron pairs and gammas); also
their optical Cerenkov emission is partially  suppressed by the
horizontal air opacity.
   However this suppression acts also as a useful filter
  leading to the selection of higher CR events. Their Cerenkov lights
  will be scattered and partially transmitted ($1.8\cdot 10^{-2}$ at $551$ nm., $6.6\cdot 10^{-4}$ at $445$ nm.)
  depending on the exact zenith angle and the seeing: assuming an average suppression factor -  $5\cdot
  10^{-3}$- and  the  nominal Magic threshold at $30$ GeV, this
      corresponds to a hadronic shower coming from the horizon
  with an  energy above $E_{CR}\simeq 6 $ PeV.
%%(diluted by nearly three order of magnitude by larger distances)
     Their primary flux may be estimated considering
     the known cosmic ray fluxes at the  same energy on the top of the atmosphere (both protons
    or helium) (see DICE Experiment referred in \cite{Grieder01}) :
     $\phi_{CR}(E_{CR} = 6\cdot 10^{15} eV)\simeq 9\cdot
     10^{-12}cm^{-2}s^{-1}$.
Within a Shower Cerenkov angle $\Delta\theta = 1^{\o}$
     at a distance  $d =167 km \cdot \sqrt{\frac{h}{2.2 km}}$
     (zenith angle $\theta \simeq 87^{\o}- 88^{\o}$)
     giving  a  shower area $ [A = \pi \cdot(\Delta\theta \cdot d)^2\simeq 2.7 \cdot 10^{11} cm^2 (\frac{d}{167 km})^2 ]$,
the consequent event rate per night  for a Magic-like telescope
with a field of view of $[\Delta\Omega =(2^o \cdot 2^o)\pi \simeq
3.82 \cdot 10^{-3} sr.]$ is  \\
%%%% ($[\Delta(t)= 4.32 \cdot 10^4s]$):
\[N_{ev}=\phi_{CR}(E= 6\cdot 10^{15} eV)\cdot A \cdot \Delta \Omega
\cdot \Delta(t) \simeq 401/12 h \]
      Thus one may foresee that   nearly every
      two minutes  a horizontal hadronic  shower  may be observed by  Magic if it were pointed towards the horizon
      at zenith angle $87^o-88^o$.  Increasing the altitude $h$ of the observer, the
      horizon zenith angle grows: $\theta \simeq [90^o + 1.5^o
      \sqrt{\frac{h}{2.2km}}]$.
       In analogy at a more distant horizontal edges (standing at height $2.2
       km$ as for Magic, while observing at zenith angle $\theta \simeq 89^o- 91^o$
         still above the horizons) the observation range $d$ increases : $d= 167\sqrt{\frac{h}{2.2 km}} + 360 km = 527
       km$;  the consequent shower area widen by more than an order of
       magnitude (and more than  three order respect to vertical showers) and the foreseen event number,
       now for much harder CR at $E_{CR} \geq 3\cdot 10^{17} eV$,  becomes:
        $$N_{ev}=\phi_{CR}(E= 3\cdot 10^{17} eV)\cdot A \cdot \Delta \Omega \cdot
      \Delta(t) \simeq 1.6 /12 h$$
       Therefore at  $\theta \simeq 91.5^o$, once per night, a UHECR around EeV energies,
      may blaze  the Magic (or Hess,Veritas, telescopes).
        A long trail of secondary muons is associated to each of these far primary Cherenkov flash  in a very huge
        area.
%%%%%%%addendum3%%%%%%%%%%%%%%
  An additional characteristic of the horizontal muon bundle are
  their common final separated  net bundle charge (all positive or
  negative); also earlier electron shower pairs are separated but effectively
  only at   highest altitudes ( few tens km.) where air density is negligible.
  These earlier shower rings  shine at wider angles some
  additional Cherenkov precursor lights.
%%%%%%%addendum3%%%%%%%%%%%%%%

          The muon bundle showering nearby the telescope, while they decay
        into electrons in flight, (source of  tens-hundred GeVs  mini-gamma showers)
        is  also detectable at a rate discussed in the following section
        (see also important earlier studies \cite{Cillis2001},\cite{NEVOD},\cite{Decor} on muon bundles at horizons).
\section{Three Muon Cerenkov Signature:  Arcs, Rings and Gamma by
$\mu^\pm \rightarrow e^\pm \rightarrow\gamma, $}
    As already noted the  photons from a horizontal UHECR  may be also revived by its secondary
    muons produced  at tens-hundred GeVs muons: they can  either  decay in flight as a gamma
    flashes,  or they may hit the telescope and their muon Cerenkov lights  "paint" arcs or rings  within the detector.
    Indeed these secondary very penetrating muon bundles    may cover distances of hundreds km  ($\simeq 600 km \cdot\frac{E_{\mu}}{100\cdot GeV}$) away from the shower origin.
    To be more precise a part of the muon primary energy will dissipate along a path in the air of $360$ km  (nearly a
      hundred GeV energy), thus a primary $130-150$ GeV muon will reach a final
       $30-50$ GeV energy, the minimal  Magic threshold value.
    Let us remind the characteristic multiplicity of secondary muons in a shower:
    $ N_\mu \simeq 3\cdot 10^5 \left( \frac{E_{CR}}{PeV}\right)^{0.85} $ \cite{Cronin2004} for GeV muons.  For the harder component (around 100
   GeV), the  muon number is  reduced almost inversely proportionally to energy
  $ N_\mu(10^2\cdot GeV) \simeq 1.3\cdot 10^4 \left( \frac{E_{CR}}{6 \cdot
  PeV}\right)^{0.85}$.
   These values must be compared to the larger peak multiplicity (but much lower energy) of
   electro-magnetic showers: $ N_{e^+ e^-} \simeq 2\cdot 10^7 \left(
   \frac{E_{CR}}{PeV}\right); N_{\gamma} \simeq  10^8 \left(\frac{E_{CR}}{PeV}\right) $.
    As mentioned before, most of the electromagnetic tail  is lost (exponentially) at
  horizons (for a slant depth larger than a few hundreds of
  $\frac{g}{cm^2}$), excluding re-born, upgoing $\tau$ air-showers
  \cite{Fargion2004},\cite{Fargion2004b} to be discussed later.
   Therefore   gamma-electron pairs are only partially  regenerated
    by the penetrating muon decay in flight, $\mu^\pm \rightarrow \gamma, e^\pm$
   as a parasite  electromagnetic showering \cite{Cillis2001},\cite{Cronin2004}.
   Indeed $\mu^\pm $  may decay in flight (let say  at $100$ GeV energy,at $2-3\%$ level within a $12-18$ km distances)
    and they may inject more and more lights, to their primary (far born) shower beam.
    The ratio between $\gamma$ over $\pm \mu$ offer a  clear hint of the
    Shower evolution \cite{Fargion2004b}.  These tens-hundred GeVs  horizontal muons and their associated mini-Cerenkov $\gamma$ showers are generated
    by either a single muon mostly produced at hundreds of kilometers  by
a single  primary   hadron  with an energy of  hundreds GeV-TeV or
rarer muons, part of a wider horizontal  bundle of large
multiplicity born at TeVs-PeV  or higher energies, as secondary of
an horizontal shower.     Between the two cases there is a smooth
link.     A whole continuous spectrum  of multiplicity  begins
from a unique muon up to a multi muon shower production.
     The  dominant noisy "single" muons at hundred-GeV energies
     will lose memory of their primary low energy and  hidden  mini-shower, (a hundreds GeV or TeVs hadrons );
      a single muon  will blaze just alone.
    The frequency of muon "single" rings or arcs  is much larger
     than the muon bundles and it is based on solid observational data
    (\cite{Iori04} ; \cite{Grieder01},\cite{NEVOD},\cite{Decor}
     as shown in Fig.2  see also the references on the  MUTRON experiment therein).
     The  event number due to these   " single noise" is:\\
     $$N_{ev- \mu} (\theta = 90^o)= \phi_{\mu}(E\simeq 10^{2} eV) \cdot A_{Magic} \cdot \Delta \Omega \cdot
      \Delta(t) \simeq 120 /12 h$$. Their muon abundance (TeV primary over PeV primary CR) is
      nearly $20$.    The additional gamma  mini-showers around the telescope due to a decay
         of those muons in flight (with a probability $p\simeq 0.02$),
         %recorded within a
         %larger collecting  Area $A_{\gamma} \geq 10^9 cm^2$
         is even a more frequent source of noise  (by a factor $\geq
         8$):\\
       $$N_{ev- \mu\rightarrow \gamma}\geq \phi_{\mu}(E\simeq 10^{2} eV)\cdot p \cdot A_{\gamma} \cdot \Delta \Omega \cdot
      \Delta(t) \simeq 960 /12 h$$
      These   single background gamma-showers must take place nearly once
       per minute (in a negligible hadronic background) and they represent a useful  tool
       to calibrate the possibility of detecting  Horizontal CR.

 \section{Muons clustering with their early Showers Blaze}
    On the contrary PeVs (or higher energy) CR shower Cerenkov lights
     maybe  observed, more (nearly twenty times) rarely, in coincidence  both by their primary and
     by their later secondary arc and gamma mini-shower.
   Their $30-100$ GeV  energetic muons are flying  nearly un-deflected
  $\Delta \theta \leq 1.6^o \cdot \frac{100 \cdot GeV}{E_{\mu}}\frac{d}{300 km}$
  for a characteristic horizons distances d , only partially bent by the geo-magnetic fields ($\sim$ $0.3$
  Gauss).  As mentioned, to flight   through the whole horizontal air column depth
  ($360$ km equivalent to $360$ water depth) the muon   lose nearly $100$ GeV;
   consequently the initial muon energy should be a little  above this threshold
   to be observed by Magic: (at least $ 130-150 $ GeV).   The deflection angle is  small:
    $\Delta \theta \leq 1^o \cdot \frac{150 \cdot GeV}{E_{\mu}}\frac{d}{300  km}$).
   Given the area of Magic ($A = 2.5 \cdot 10^6 cm^2$) we expect   roughly the
   following number of events due to  direct muons hitting the Telescope, flashing  as rings and arcs, each
   night:\\
  $$N_{ev}=\phi_{CR}(E= 6\cdot 10^{15} eV)\cdot N_\mu(10^2\cdot GeV) \cdot A_{Magic} \cdot \Delta \Omega \cdot
      \Delta(t) \simeq 45 /12 h$$ \\
to be correlated (at $\simeq 10 \%$ probability or less) with the
above results of $401$ primary Cerenkov flashes at the far
distances. Moreover, the same muons are decaying in flight  at a
minimal probability $2\%$   leading to   mini-gamma-showers  in a
wider area ($A_{\gamma}= 10^9 cm$). The related mini-gamma shower
number of events we expect
   is:\\
   $$N_{ev}= \phi_{CR}(E= 6\cdot 10^{15} eV)\cdot N_\mu(10^2\cdot GeV) \cdot p \cdot A_{\gamma} \cdot \Delta \Omega \cdot
      \Delta(t) \simeq 360 /12 h$$\\      Therefore,  at $87^{\o}-88^{\o}$ zenith angle, there is a flow
      of  dozens event a night of    primary CR (at $E_{CR}\simeq 6\cdot 10^{15} eV$), whose earliest
      showers, consequent secondary muon-arcs as well as   nearby muon-electron mini-shower
         take place at comparable rate (one every  $120-600$ in twin correlation or one an hour for multi-correlation ).
          These $certain$ and frequent clustered signals offer
          an unique tool for    calibrating Magic (as well as  Hess,Cangaroo,Veritas Cerenkov Telescope Arrays)
           for Horizontal High Energy Cosmic Ray Showers. Rarer events ( a dozen a night)
            may contain at once both Rings,Arcs       and tail    of gamma  shower and twin dots
             at a few ($1-2$) degree separation (due to geomagnetic splitting) by
             Cerenkov primary  forked shower.    It is possible to estimate also
       the observable muon-electron-Cerenkov  photons  from up-going  albedo muons observed by the most
        recent ground experiments  \cite{NEVOD}  \cite{Decor}:
         their flux   is already suppressed at zenith angle $91^o$ by at least two orders of
         magnitude and by four orders for up-going zenith  angles $94^o$.
   Pairs or bundles are nevertheless rarer
   (up to $\phi_{\mu} \leq 3 \cdot 10^{-13} cm^{-2}s^{-1}sr^{-1}$   \cite{NEVOD}  \cite{Decor}).
   They are never associated to up-going showers
    (excluding the case of   tau air-showers or  the Glashow $\bar{\nu_e}-e\rightarrow W^-$ and  the comparable $\chi^o + e\rightarrow
   \tilde{e}$) and not competitive to up-going tau secondaries discussed below.
%%%   detectable by stereoscopic Magic or Hess array telescopes, selecting and evaluating their column depth origination, just discussed below.
%---------------------------------------------
\section{UHE $\bar{\nu_e}-e\rightarrow W^-$ and $\chi^o + e\rightarrow \tilde{e}$ resonances versus $\tau$ air-showers }

  The appearance of horizontal UHE
   $\bar{\nu_\tau}$ ${\nu_\tau}\rightarrow \tau$ air-showers (Hortaus or Earth-Skimming neutrinos)
    has been widely studied \cite{Fargion1999},\cite{Fargion
2002a},\cite{Bertou2002},\cite{Feng2002}; see also
\cite{Fargion03},\cite{Fargion2004},\cite{Jones04},
  \cite{Yoshida2004},\cite{Tseng03}, \cite{Cao} and more recent \cite{Fargion2004b}.
    Their rise from the Earth is source of rare clear signals for neutrino UHE astronomy (see fig.3).
   However also horizontal  events by UHE $6.3$ PeV, Glashow $\bar{\nu_e}-e\rightarrow W^-$ and a possible
   comparable SUSY
     $\chi^o + e\rightarrow \tilde{e}$ \cite{Datta04} hitting and showering in air have a non negligible number of events;
   one should remember that at peak resonance the probability conversion is $3 \cdot 10^{-3}$ at 150 km air distance:
   $$N_{ev}= \phi_{\bar{\nu_e}} (E= 6\cdot 10^{15} eV) \cdot A\cdot \Delta \Omega \cdot
      \Delta(t) \simeq  5.2 \cdot 10^{-4}/12 h$$
      assuming the minimal  GZK neutrino flux : $\phi_{\bar{\nu_e}} (E= 6\cdot 10^{15} eV)\simeq 5 \cdot
      10^{-15}$ eV $ cm^{-2} s^{-1}sr^{-1}$.      Therefore   in a year of observations, provided that the data are taken at  night, assuming a minimal GZK flux,  a crown array of
   a $90$ Magic-like telescopes set on a circle  ($2\cdot \pi = 360^o$) and facing the horizon,
   would give a number of events  comparable to a   $Km^3$ detector, (nearly a dozen per year). Indeed  Magic pointing
   at the horizon  offers a detection comparable to   the present AMANDA $\simeq  1\%  Km^3$ effective volume.
     A better efficency (but not higher rate)  occurs at EeV energies for tau (by GZK neutrino) induced air-showers \cite{Fargion2005}.
     The same result may be improved assuming the coexistence of a
      scintillator Crown Array able to verify the electromagnetic
      and muon shower in coincident arrival time.  The result may be
      greately increased by a number of multiple crown array
      located at different (km) distances and altitudes.
          The Crown Area may encompass a large area
       as hundreds square meters array (see last figure) and
      its structure may be added around the Cerenkov Crown telescope
      array at the same mountain top. Its ability to discover muons (sometimes in correlation with the Cerenkov telescope)
      as well as the electromagnetic trace $\gamma, e^+,e^-$ nature, for a total area of
      63 $m^2$, is superior by a factor two respect a  (single ) Magic telescope, because it  enjoys a
      larger solid angle and a longer  recording time. Magic-Crown
      and Muon Crown array may reveal tens of thousands horizontal
      CR shower and a ten of up-going neutrino induced air-shower
      a year (for GZK minimal fluxes).   To conclude, while Magic looks
        upward to investigate the Low Gamma GeVs Astronomy,
    the same telescope looking at the horizon may well see higher
    (PeVs-EeVs) CR, and rarely along the  edge,
     GZK $\bar{\nu_e}-e\rightarrow W^-$ neutrinos; finally , below the horizons
    $\nu{\tau} \rightarrow \tau$ air-showers arise and, surprisingly
   even  SUSY $\tilde{e}$ lights may come in the sky (with electronmagnetic
   showers).  Finally the possible detection of up-going few-tens GeV muons
   (induced by   energetic Solar Flare) at a rate $\phi_{\mu}\simeq 10^{-10}
   cm^{-2}s^{-1}$,for few minutes, during the maximal solar flare activity,
    might also be a target of such New Muon-Cerenkov astronomy \cite{Solar}.

 \begin{figure}[h]
\begin{center}
\includegraphics[width=14cm]{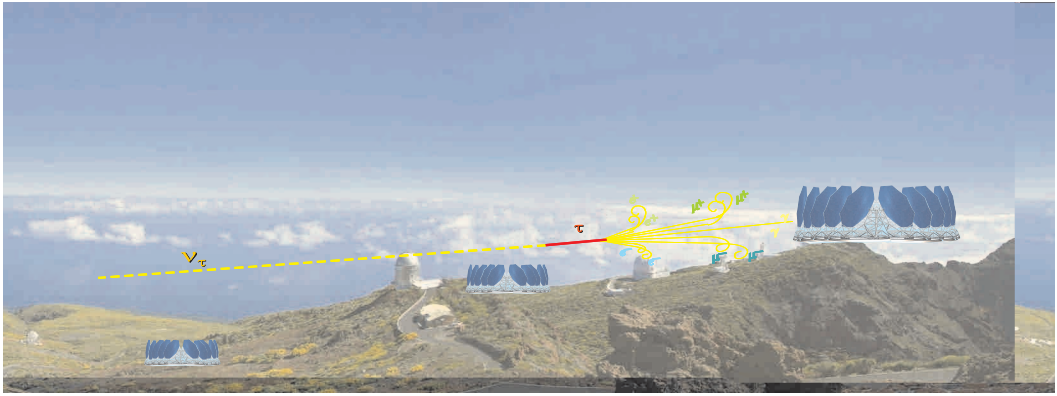} \end{center}
\vspace{0.3cm} \caption{The Earth view from Canarie sites for a
Magi-like Crown Telescope Array facing the Horizons, blazed, in
dark nights, by CR showers and rare up-going Tau Air-Shower
Cerenkov flashes; the  telescopes in circular array (of tens
telescopes at $360^o$ ) may test a wide area volume, nearly a
$km^3$ of air mass. The present (real) single Magic telescope is
located in the bottom left corner of the picture }
%\vspace{0.1cm} \label{figpbar} \end{figure}
\end{figure}

 \begin{figure}[t]
\begin{center} %\vspace{-1.0cm}
\includegraphics[width=10cm]{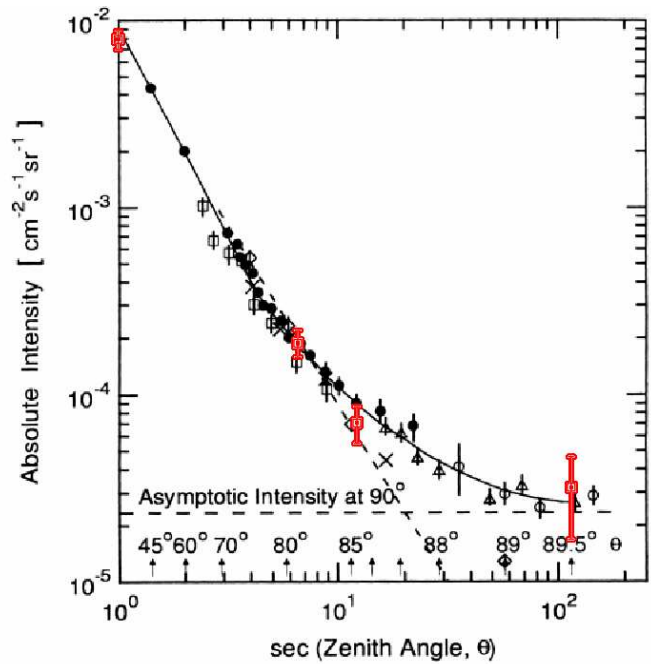}
\vspace{-0.3cm} \caption{Muon fluxes at different zenith angle
from observational data \cite{Iori04}, \cite{Grieder01} }
\end{center}
\end{figure}

 \begin{figure}[t]
\begin{center} %\vspace{-1.0cm}
\includegraphics[width=12cm,height=7cm]{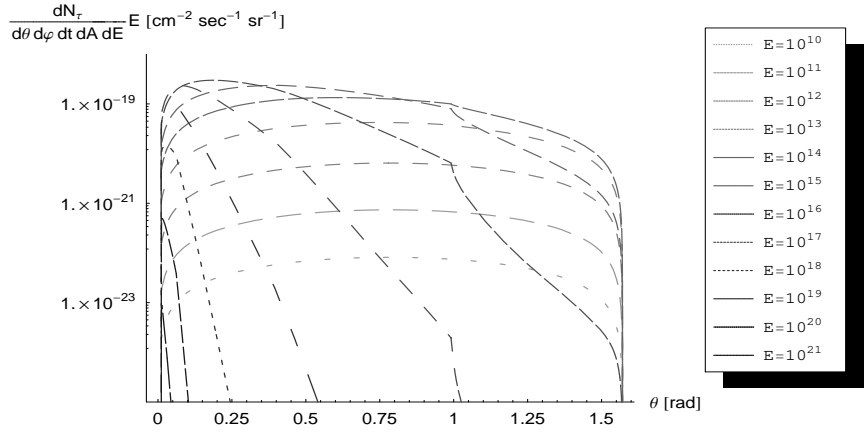} \end{center}
\vspace{-0.3cm} \caption{Differential Tau Number flux by GZK
neutrino flux and consequent Hortau Earth Skimming flux rate from
the Earth }
%\vspace{0.1cm} \label{figpbar} \end{figure}
\end{figure}

 \begin{figure}[t]
\begin{center} %\vspace{-1.0cm}
\includegraphics[width=13cm,height=7cm]{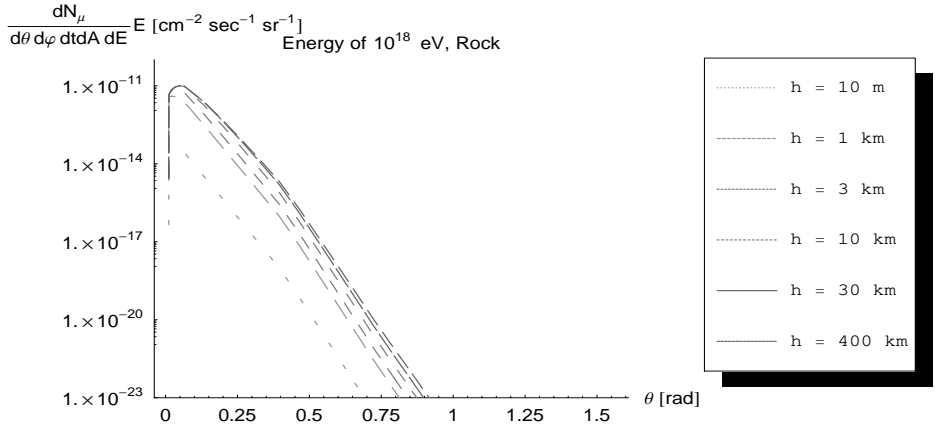} \end{center}
\vspace{-0.3cm} \caption{Consequent Differential Muon Number flux
by GZK neutrino and Tau flux  Hortau Earth Skimming the Earth,
and consequent secondary muons: because a different acceptance at
different height the flux depend on the observer quota}
%\vspace{0.1cm} \label{figpbar} \end{figure}
\end{figure}

\begin{figure}[t]
\begin{center} %\vspace{-1.0cm}
\includegraphics[width=12cm,height=12cm]{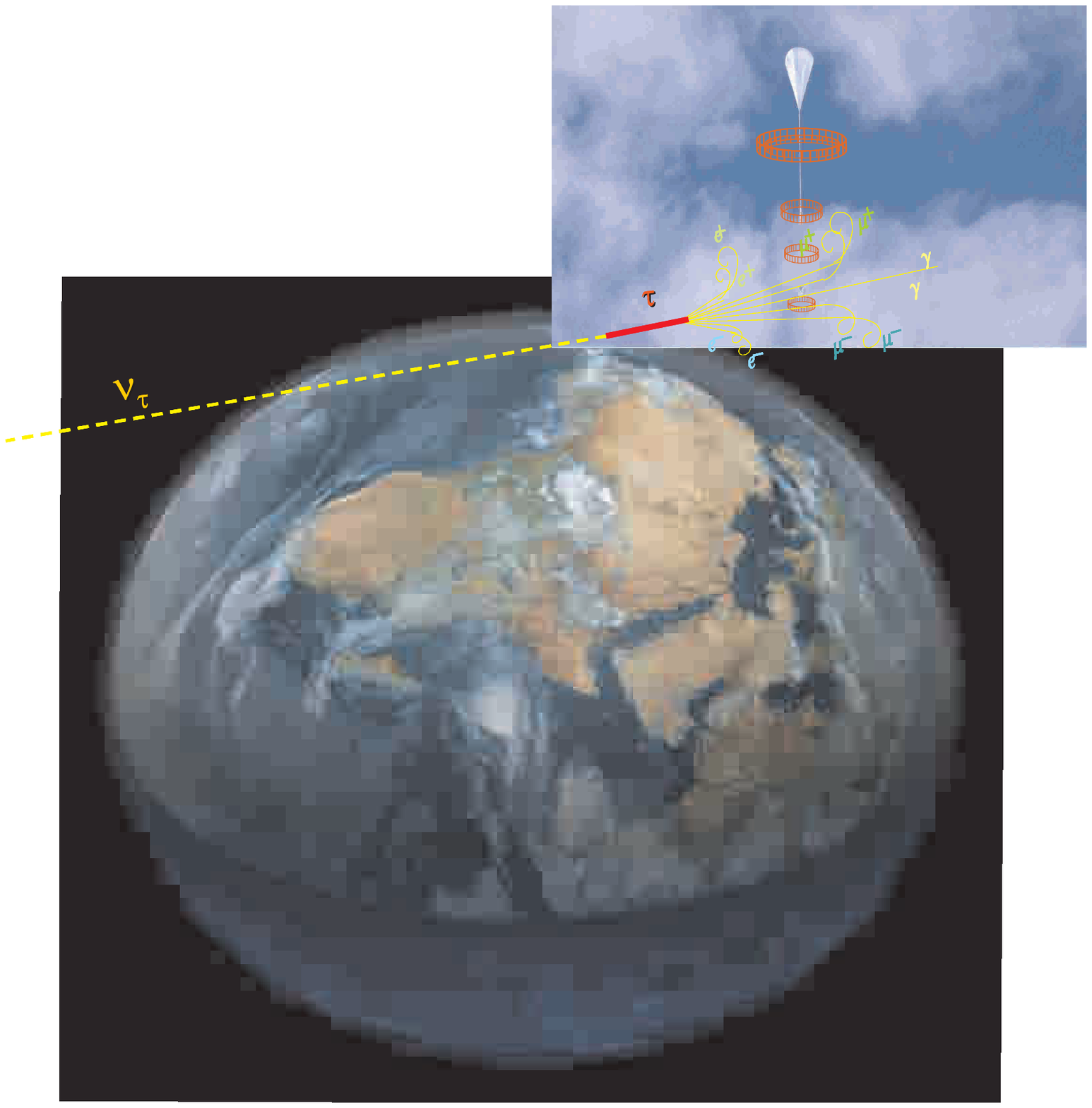}
\vspace{0.3cm} \caption{ A schematic figure of a Crown circular
detector array in Balloon (or located on the top of a mountain) by
a $\simeq 10$ meter radius size, whose  inner-outer aligned tiles
are able to reveal, by time of flight, the crossing and the
azimuth directions of horizontal muons (and electron pairs and
gamma) bundles; in the picture one  an upward  Horizontal Tau
Air-Shower (label by $\tau$); the  $360^o$ tile disposal
guarantees a wide azimuth solid angle view. The twin (upper and
lower) rings distance (tens- or hundred meters) is able to better
disentangle the zenithal arrival direction of the Horizontal
Shower particles above or below the edges }
%\vspace{0.1cm} \label{figpbar} \end{figure}
\end{center}
\end{figure}

%\label{}
% \appendix

% \section{}
% \label{}

% Bibliographic references with the natbib package:
% Parenthetical: \citep{Bai92} produces (Bailyn 1992).
% Textual: \citet{Bai95} produces Bailyn et al. (1995).
% An affix and part of a reference:
%   \citep[e.g.][Ch. 2]{Bar76}
%   produces (e.g. Barnes et al. 1976, Ch. 2).


\begin{thebibliography}{100}



%%%\bibitem[Datta et al. 2004]{Datta04}{\normalsize  Datta A.,  Fargion D., Mele B; hep-ph/0410176}
\bibitem{Datta04}{\normalsize  Datta A.,  Fargion D., Mele B; hep-ph/0410176; JHEP 0509 (2005) 007}


\bibitem{Bertou2002}  {\normalsize  Bertou, X. et al.
2002, Astropart. Phys.,  17,  183}


%
\bibitem{Cao} Z. Cao, M. A. Huang, P. Sokolsky, Y. Hu, astro-ph/0411677.

\bibitem{Cillis2001}{\normalsize Cillis, A.N.,  \& Sciutto, S.J., 2001, Phys. Rev. D64, 013010}

\bibitem{Cronin2004}{\normalsize Cronin, J.W., 2004, TauP Proceedings, Seattle 2003; astro-ph/0402487}

%%%%\bibitem{Fargion-Mele-Salis99}
%%%%%%%%%%{\normalsize Fargion, D., Mele, B., \& Salis, A., 1999, ApJ 517,
%%%%%%%725; astro-ph/9710029 }

\bibitem{Fargion1999}  {\normalsize Fargion, D., Aiello,et.all. 1999, 26th ICRC
HE6.1.10,396-398 }

%%%%%\bibitem[Fargion 2002]{Fargion 2002a}  {\normalsize Fargion, D., 2002, ApJ, 570, 909; see astro-ph/0002453/9704205 }
\bibitem{Fargion 2002a}  {\normalsize Fargion, D., 2002, ApJ, 570, 909; see astro-ph/0002453/9704205 }



%%%%\bibitem[Fargion et al. 2003]{Fargion03}  {\normalsize
%%%%Fargion, D. et al. 2003, Recent Res. Devel.Astrophysics., 1, 395; astro-ph/0303233} .

\bibitem{Fargion03}  {\normalsize
Fargion, D. et al. 2003, Recent Res. Devel.Astrophysics., 1, 395;
astro-ph/0303233} .
%

%%%%\bibitem[Fargion et al. 2004a]{Fargion2004}{\normalsize Fargion D., De Sanctis Lucentini, P.G., De Santis, M., Grossi, M.,
%%%% 2004, ApJ, 613, 1285; hep-ph/0305128.}
\bibitem{Fargion2004}{\normalsize Fargion D., De Sanctis Lucentini, P.G., De Santis, M., Grossi, M.,
 2004, ApJ, 613, 1285; hep-ph/0305128.}

%%%%\bibitem[Fargion et al. 2004b]{Fargion2004b}{\normalsize  Fargion D. et al., Nuclear Physics B (Proc. Suppl.), 136 ,119(2004) }
\bibitem{Fargion2004b}{\normalsize  Fargion D. et al., Nuclear Physics B (Proc. Suppl.), 136 ,119(2004) }
\bibitem{Fargion 2004c}{\normalsize  Fargion D. astro-ph/0412582, Frontier Science, Phys. and Astrophysics in Space (2004) }
\bibitem{Fargion2005}{\normalsize  Fargion D. astro-ph/0511597, in Prog.Part. Nucl. Phys.(2005) }


%%%%\bibitem[Feng 2002]{Feng2002}  {\normalsize Feng, J.L.,
%%%%Fisher, P., Wilczek, F., \& Yu, T.M., 2002, Phys. Rev. Lett. 88,
%%%%161102; hep-ph/0105067 }
\bibitem{Feng2002}  {\normalsize Feng, J.L.,
Fisher, P., Wilczek, F., \& Yu, T.M., 2002, Phys. Rev. Lett. 88,
161102; hep-ph/0105067 }

%%%%%%%%%\bibitem{Gandhi96}{\normalsize Gandhi, R., Quigg, C., Reno, M.H., Sarcevic, I., 1996, Astrop. Phys., 5,  81}

%%%%%%\bibitem[Gandhi et al. 1998]{Gandhi98}{\normalsize Gandhi, R., Quigg, C., Reno, M.H., Sarcevic, I., 1998, Phys. Rev. D., 58, 093009}
\bibitem{Gandhi98}{\normalsize Gandhi, R., Quigg, C., Reno, M.H., Sarcevic, I., 1998, Phys. Rev. D., 58, 093009}

%
%%%%%\bibitem[Grieder 2001]{Grieder01}{\normalsize P.K.F. Grieder , Cosmic Rays at Earth, Elsevier 2001 }
\bibitem{Grieder01}{\normalsize P.K.F. Grieder , Cosmic Rays at Earth, Elsevier 2001 }

%%%%%%\bibitem[Iori et al. 2004]{Iori04}{\normalsize Iori, M., Sergi, A., Fargion, D., 2004, astro-ph/0409159}
\bibitem{Iori04}{\normalsize Iori, M., Sergi, A., Fargion, D., 2004, astro-ph/0409159, astro-ph/0602108}

%%%%%%\bibitem[Jones et al. 2004]{Jones04}{\normalsize Jones, J. et al.
%%%%%%%%%%%%%%%%%Mocioni, I., Reno, M.H., Sarcevic, I.,
%%%%%%2004  Phys. Rev. D , 69, 033004}

\bibitem{Jones04}{\normalsize Jones, J. et al.
%%%%%%%%%%%Mocioni, I., Reno, M.H., Sarcevic, I.,
2004  Phys. Rev. D , 69, 033004}



%%%%%%%%%\bibitem{Learned Pakvasa 1995}  {\normalsize Learned, J.G., \& Pakvasa, S., 1995, Astropart. Phys., 3, 267 }
%

%%%%%%%%\bibitem{Matthews01}{\normalsize J.C. Matthews (2001), Proc. 27th ICRC (Hamburg), 1, 161.}


\bibitem{Sine2001}
{\normalsize Sinegovskaya T.S. and Sinegovsky S.I.2001, Phys.Rev.
D63, 096004}
%%%%%%\bibitem[Tseng et al. 2003]{Tseng03}
%%%%%%{\normalsize Tseng, J.J, Yeh, T.W., Athar et al.
%%%%%% %%%%%%H., Huang, M.A., Lee, F.F., Lin, G.L.,
%%%%%%  2003, Phys. Rev. D68, 063003}

\bibitem{Tseng03}
{\normalsize Tseng, J.J, Yeh, T.W., Athar et al.
 %%%%%%H., Huang, M.A., Lee, F.F., Lin, G.L.,
  2003, Phys. Rev. D68, 063003}



%
%%%%%%\bibitem[Yahin et al. 2003]{NEVOD}{\normalsize I.I. Yashin et al., ICRC28 (2003), 1195.}
\bibitem{NEVOD}{\normalsize I.I. Yashin et al., ICRC28 (2003), 1195.}

%%%%%%%\bibitem[Yashin et al. 2003]{Decor}{\normalsize I.I. Yashin et al, ICRC28 (2003), 1147.}
\bibitem{Decor}{\normalsize I.I. Yashin et al, ICRC28 (2003), 1147.}



%%%%%%%\bibitem[Yoshida et al. 2004]{Yoshida2004}{\normalsize  Yoshida, S.,et al
%%%%%%%%%%%%%%Ishibashi, R., Miyamoto, H.,
%%%%%%%2004, Phys. Rev. D69, 103004};
%%%%%%%%%astro-ph/0312078}
%

\bibitem{Yoshida2004}{\normalsize  Yoshida, S.,et al
%%%%%%%Ishibashi, R., Miyamoto, H.,
2004, Phys. Rev. D69, 103004};

\bibitem{Solar} D.Fargion,F.Moscato,Chin. J. Astron. Astrophys.Vol 3.Suppl.75-86. 2003.




\end{thebibliography}
\end{document}